\documentclass[twocolumn,showpacs,preprintnumbers,amsmath,amssymb]{revtex4}

\usepackage{graphicx}
\usepackage{dcolumn}
\usepackage{bm}

\begin{document}


\title{An EOM-assisted wave-vector-resolving Brillouin light scattering setup}

\author{T. Neumann}
 \email{neumannt@physik.uni-kl.de}
\author{T. Schneider}
\author{A. A. Serga}
\author{B. Hillebrands}
\affiliation{%
Fachbereich Physik and Forschungszentrum OPTIMAS\\
Technische Universit\"at Kaiserslautern, 67663 Kaiserslautern, Germany}%

\date{\today}

\begin{abstract}
Brillouin light scattering spectroscopy is a powerful technique which incorporates several extensions
such as space-, time-, phase- and wave-vector resolution.
Here, we report on the improvement of the wave-vector resolution by including an electro-optical
modulator. This provides a reference to calibrate the position of the diaphragm hole which is used for
wave-vector selection. The accuracy of this calibration is only limited by the accuracy of the
wave-vector measurement itself. To demonstrate the validity of the approach the wave vectors of
dipole-dominated spin waves excited by a microstrip antenna were measured.
\end{abstract}

\maketitle

%

\section{Introduction}
Brillouin light scattering (BLS) spectroscopy is a versatile technique to investigate dynamic magnetic
phenomena. Its power is significantly increased by numerous extensions which were added to the basic
spectroscopic setup over time.

By introducing time-resolution, it was possible to investigate the evolution of a parametrically excited
magnon gas in a ferrite film and observe the formation of a Bose-Einstein condensate of magnons at room
temperature \cite{Dem06, Dem08}. The inclusion of space-resolution led to the discovery of important
nonlinear wave phenomena such as soliton and bullet formation \cite{Bau98, Ser08}, the observation of
spin-wave tunneling \cite{Dem04} as well as spin-wave quantisation in nanoscaled structures
\cite{Schu08}.

Other discussed extensions comprise phase- \cite{Ser06} and wave-vector resolution \cite{Xia98}. The
latter is particularly interesting since the frequency does usually not uniquely identify a wave
eigenmode. Due to an often complex dispersion relation the additional knowledge of the wave vector is
essential.

For the investigation of spin waves in ferrite films by BLS, wave-vector resolution was already
introduced in the late 70's \cite{Ven79}. Subsequently, many studies (e.g. \cite{Wet83, Kab94, Kab97})
have copied the originally presented principle: to the BLS setup in forward scattering geometry a
diaphragm is added  in the beam path after the sample stage (see Fig.~\ref{fig:aufbau}). Depending on
the shape and position of the hole in the diaphragm, some components of the scattered laser beam which
correspond to certain in-plane wave vectors are blocked while others can pass and are detected. The
latest success of this technique was the time- and wave-vector resolved observation of a parametrically
pumped magnon gas after pumping was switched off \cite{Dem08a}.

It is of crucial importance for an accurate measurement of the in-plane wave-vector to calibrate the
position of the diaphragm hole. The zero position when the elastically scattered beam passes through the
diaphragm can be adjusted by sight, however this procedure is inevitably inaccurate. A second
possibility is to calibrate the position based on the measured data: Stokes and anti-Stokes peaks in the
BLS spectrum lead to two distinct signals with opposing wave vectors. Their symmetric position with
respect to the center can in special cases be used for calibration. However, the intensities of Stokes
and anti-Stokes peaks can differ greatly which makes this procedure difficult. Moreover, it is
unnecessarily time-consuming since the weaker of the two signal peaks determines the accuracy of the
calibration and, therefore, the required measurement time though in many cases it does not yield any
additional physical information about the system under investigation.

The approach presented here uses an intrinsic calibration which is achieved by placing an
electro-optical modulator (EOM) in the optical path behind the laser light source. The small amount of
modulated, frequency-shifted light plays the role of a reference beam for the wave-vector resolution.
Since it follows the same path as the unshifted laser light scattered inelastically from the sample but
does (in first approximation) not undergo any inelastic scattering in the sample itself, it indicates
the position where the in-plane wave vector vanishes for the measurement.

The proposed calibration procedure has two major advantages. First of all, it is applicable even when
one of the two signals form Stokes and anti-Stokes scattering is not large enough to be observed. This
is in particular the case for surface magnetostatic spin waves. Secondly, it potentially decreases the
measurement time since the wave vector scanning does not have to be performed over the whole range of
wave vectors but can (if a symmetry is already known) be restricted to one of the symmetric parts.

To test the validity of the presented method the in-plane wave vectors of propagating, dipole-dominated
spin waves were resolved. The obtained results are in good quantitative agreement with theory.

It should be remarked that electro-optical modulators have already been used to realize phase resolution
\cite{Ser06} and enhance the frequency resolution \cite{Cap01}. This work adds to their increasing role
for the improvement of the Brillouin light scattering setup.

\section{Setup}
\begin{figure}[t]
\includegraphics[height = 32ex]{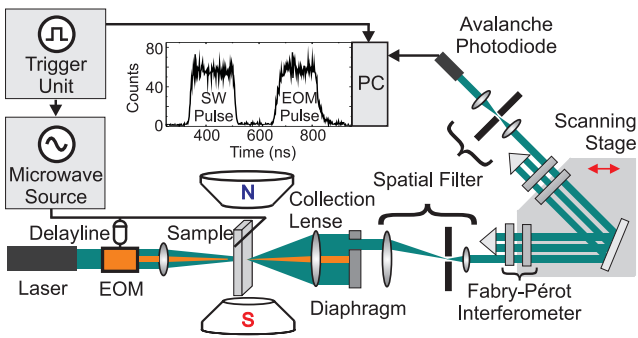}
\caption{\label{fig:aufbau} (Color online) Sketch of the experimental setup.}
\end{figure}

The experimental setup is shown in Fig.~\ref{fig:aufbau}. The sample under investigation consisted of a
$5~\mu{\rm m}$ thick yttrium-iron-garnet (YIG) film which was tangentially magnetized by a magnetic
field $H$ (indicated by two pole pieces in Fig.~\ref{fig:aufbau}). To a microstrip transducer on the
surface of the YIG stripe a $200~{\rm ns}$ long microwave pulse with $7.132~{\rm GHz}$ carrier frequency
was supplied with a $1~\mu{\rm s}$ repetition rate. The configuration was chosen in such a way that the
microwave pulse excites a packet of backward volume magnetostatic spin waves (BVMSW) which propagates in
the film in the direction of the bias magnetic field \cite{Dam61, Hur95}. Thus, the in-plane wave vector
of the excited spin waves has a well defined, unique non-zero component. The measurements discussed
below focus on determining the wave number of this wave.

To detect the spin-wave packet light from a single mode, frequency-stabilized $532~{\rm nm}$ laser was
focused on the sample close to the antenna. The transmitted light was sent to a (3+3)-pass tandem
Fabry-P\'erot interferometer where the frequency of the light inelastically scattered from the spin
waves was resolved. A good description of the underlying BLS setup is found in \cite{Hil99}.

The existing BLS setup already includes time- and space resolution in the following way: In order to
probe different points of the sample, the sample is mounted on a stage which can be moved by a stepper
motor. Time resolution is achieved by measuring the time between the launch of the microwave signal
pulse which excites the spin-wave packet and the detection of the scattered photons by the detector. In
the current setup the time resolution is limited to $1.8~{\rm ns}$ due to the finesse of the
Fabry-P\'erot etalons. A detailed account is given in \cite{Bue00}.

Wave-vector resolution was added to the existing setup by placing a diaphragm with a central hole of
$0.5~{\rm mm}$ diameter in the focal plane behind the collection lens. The diaphragm was mounted on a
stage which was horizontally movable by a PC-controlled stepper motor. The stage was moved in steps of
size $0.08~{\rm mm}$. Since the investigated spin waves possessed only one non-vanishing in-plane
component of the wave-vector the chosen one-dimensional approach is sufficient for demonstration. To
measure both in-plane wave-vector components an additional stage for the vertical displacement of the
diaphragm will be added. The measurement principle remains, however, unchanged.

To calibrate the diaphragm position, an EOM was placed in the beam path in front of the sample (see
Fig.~\ref{fig:aufbau}). It was driven by a $200~{\rm ns}$ long pulse from the same microwave source that
generated the spin-wave pulse. However, the EOM-pulse was delayed compared to the spin-wave pulse in
order to make EOM and spin-wave signal clearly distinguishable in the time-resolved measurements.

Since the same microwave frequency is applied to the EOM and the microstrip transducer which excites the
spin waves, the resulting signal peaks in the BLS spectrum coincide. This has two practical advantages.
First of all, the frequency interval which is effectively scanned by the interferometer can be small.
This reduces the overall measurement time which is particularly important for wave-vector resolved
measurements. Secondly, the EOM-signal can be used as a frequency reference \cite{Cap01}.

%

\section{Experimental results}

Figure~\ref{fig:measurement} shows the intensity of the detected BLS signal relative to the elapsed time
and the displacement of the diaphragm from its initial (arbitrary) position. Three signals are clearly
distinguishable:

\begin{figure}[b]
\includegraphics[height = 33ex]{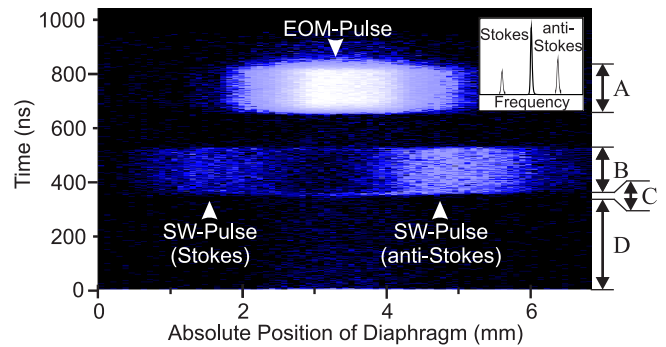}
\caption{\label{fig:measurement} (Color online) Time-resolved intensity of the scattered light with
respect to the displacement of the diaphragm for a bias magnetic field $H=4\pi~1.870~{\rm Am}^{-1}$ and
a spin-wave carrier frequency $f = 7.132~{\rm GHz}$. Indicated are the signals from EOM reference pulse
in the time interval A as well as from the spin-wave (SW) packet corresponding to the Stokes- and
anti-Stokes-peak in the BLS spectrum (sketched in inset) observed in the time interval B. The thermal
signal received during the time interval D together with the signal measured during the transition
period C is shown in the different panels of Fig.~\ref{fig:fourier}.}
\end{figure}

In the time interval marked in Fig.~\ref{fig:measurement} as $A$ the signal from the EOM-pulse is seen.
The time interval $B$ contains two signals which both stem from the propagating spin-wave packet. They
correspond to the Stokes and anti-Stokes peaks in the BLS spectrum. This was checked by restricting the
BLS measurement once to the Stokes and once to the anti-Stokes peak (see inset in
Fig.~\ref{fig:measurement}). In particular, the position of the signals corresponding to Stokes and
anti-Stokes BLS peaks exchanged their positions when the laser beam was focused on the other side of the
exciting microstrip antenna where the spin-wave packet travels in the opposite direction and the
spin-wave wave vector, therefore, changes sign.

\begin{figure}[t]
\includegraphics[height = 26ex]{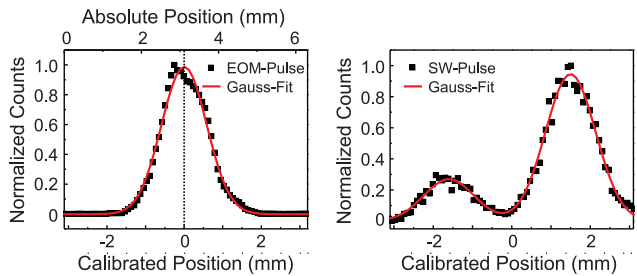}
\caption{\label{fig:shapes} (Color online) Dependence of the BLS signal obtained for the EOM pulse and
the SW pulse on the diaphragm displacement. The lines are the results of Gauss fits.}
\end{figure}

To increase the signal to noise ratio the received counts were integrated over the time intervals $A$
and $B$, respectively. The resulting intensity distribution, which depends only on the diaphragm
displacement, is shown in Fig.~\ref{fig:shapes}. It is relatively wide because of the comparatively
large pin hole in the diaphragm. However, this trade-off was accepted to decrease the measurement time.
By fitting the experimental data with a single Gaussian distribution for the EOM-signal and two
independent Gaussian distributions with the same variance for the spin-wave signal, the accuracy of the
measurement was enhanced. As can be seen from Fig.~\ref{fig:shapes}, the fits agree well with the
experimental data. The center of the Gaussian profile which fits the EOM-signal was used to calibrate
the diaphragm displacement and obtain the $k=0\,-\,$position for the diaphragm. Relative to this
position the deflection $x$ of the beam which was inelastically scattered on the spin-wave packet was
determined.
\begin{figure}[t]
\includegraphics[height = 76ex]{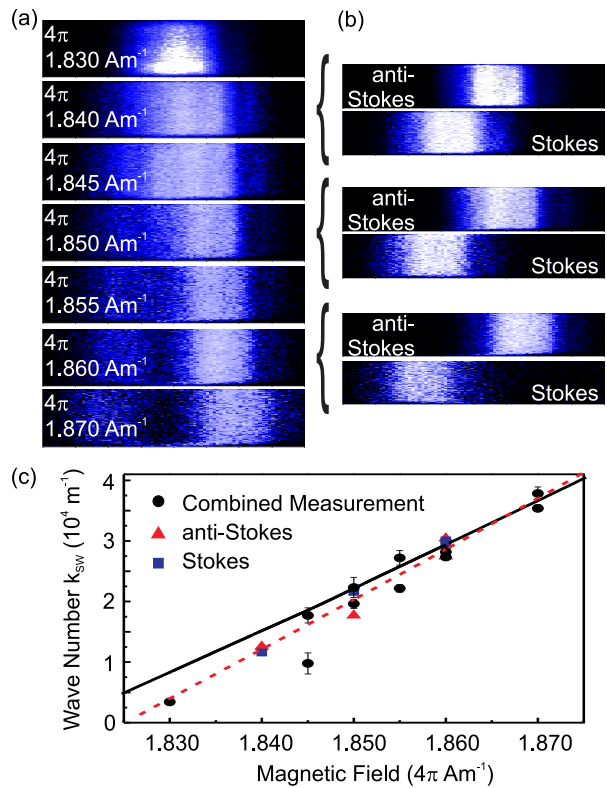}
\caption{\label{fig:field} (Color online) Dependence of the observed signal on the bias magnetic field
indicated in each panel. (a) Spatial signal distribution when Stokes and anti-Stokes are measured
together (compare to Fig.~\ref{fig:measurement}). (b) Separately measured Stokes and anti-Stokes
signals. (c) Experimentally obtained wave numbers in comparison with theoretical calculations without
fit parameters (solid line) and with adjusted magnetic field and film thickness (dashed line).}
\end{figure}

Measurements were performed for different magnetic fields with the same spin-wave carrier frequency
$7.132~{\rm GHz}$. The results are presented in Fig.~\ref{fig:field}. Figure~\ref{fig:field}(a) shows
extracts corresponding to the interval $B$ in Fig.~\ref{fig:measurement} which contains the information
on the spin-wave wave vector. In accordance with theory an almost linear field dependence is seen
\cite{Hur95}.

In order to unambiguously resolve the signals corresponding to the Stokes and anti-Stokes peaks in the
BLS spectrum, separate measurements were conducted by limiting the spectral scanning to one of the two
spectral positions (Panel (b)).

From the data, the spin-wave wave number $k_{\rm SW}$ is obtained using the Bragg-condition
\begin{displaymath}
k_{\rm SW} = 2\cdot k_{\rm Laser}\sin\big(\vartheta /2\big) =  2\cdot k_{\rm
Laser}\sin\big(\frac{\arctan(x/f)}{2}\big)
\end{displaymath}
where $k_{\rm Laser}=1.181\times10^7~{\rm m}^{-1}$ is the wave number of the incoming laser light and
$\vartheta$ is the angle between the elastically and inelastically scattered light which is determined
by the focal length $f = 500~{\rm mm}$ of the collection length and the measured deflection $x$ of the
spin-wave signal. The experimentally found spin-wave numbers $k_{\rm SW}$ are combined in
Fig.~\ref{fig:field}(c) with theoretical calculations based on the Damon-Eshbach formula for the lowest
order BVMSW mode \cite{Hur95}. The solid line has been calculated based on the measured field value, a
film thickness of $5~\mu{\rm m}$ and a saturation magnetisation $4\pi\mu_0 M_{\rm s} = 0.175~{\rm T}$.
In comparison, the dashed line is the result of a fit, where the film thickness and the magnetic field
were taken as fit parameters. The optimal value found for the thickness was $4.2~\mu{\rm m}$, the
magnetic field was adjusted by a shift $\Delta H = - 4\pi \cdot 0.007~{\rm A m}^{-1}$ relative to the
experimentally measured field $H$. Both deviations are within reasonable range. The film thickness is
not known with sufficient accuracy and is in general assumed as a fit parameter. The experimentally
measured magnetic field does not take into account any contributions from the crystalline anisotropy.

Overall, the theoretical curves agree well with experiment. The measurements indicate that the film
thickness at the point of the laser focus was less than the nominal $5~\mu{\rm m}$.


\section{Discussion}

The presented results confirm the validity of the EOM-assisted wave-vector resolution measurement
procedure. The EOM allows an easy, intrinsic calibration with the same resolution as the actual
measurement. The calibration does not rely on any symmetry in the observed peaks and can be performed
even when the scanning is restricted to one side of the BLS spectrum. The widely adjustable intensity of
the EOM beam guarantees a minimal expenditure of time to obtain a large enough signal for the analysis.

In principle, the EOM beam can be used to calibrate the diaphragm prior to the experiment. Instead, in
the presented work the EOM reference was applied parallel to the actual measurement. This is a natural
solution whenever the pulse regime is required because of other experimental restrictions. The reference
EOM beam is applied during the dead time of the cycle so that the overall duration of the experiment is
not increased and measurement as well as calibration are completed in a single run.

\begin{figure}[t]
\includegraphics[height = 66ex]{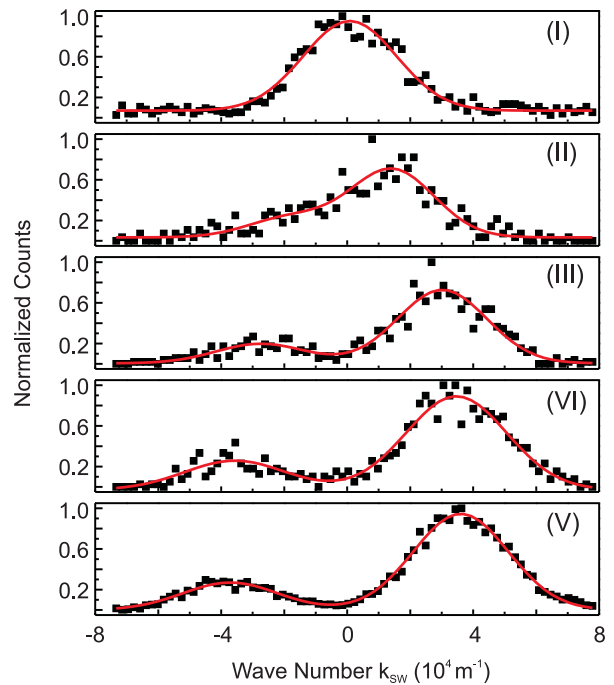}
\caption{\label{fig:fourier} (Color online) Time-resolved wave-number measurements ($\blacksquare$)
together with Gaussian fits (solid line). (I) Thermal signal from interval $D$ in
Fig.~\ref{fig:measurement}. (II)-(IV) Consecutive measurements when the pulse front passes the laser
spot (see $C$ in Fig.~\ref{fig:measurement}). (V) Spin-wave signal from $B$ in
Fig.~\ref{fig:measurement}}
\end{figure}

The experiments confirmed the applicability of the diaphragm-based approach to wave-vector resolution
for the measurement of small wave numbers. The method is not the only way to go in this regime. It is
also possible to measure the spin-wave wavelength by using phase resolution \cite{Ser06}. However, this
method relies on the scanning of the sample which does, in terms of measurement time not yield any
advantages. Moreover, it is only applicable in the case of a single spin wave propagating under
homogeneous external conditions - fast temporal or spatial variations of the wave number cannot be
resolved. The same draw-backs apply to other interference-based methods using, for instance, inductive
probes.

For the diaphragm-approach these limitations do not apply: Since the method relies on the measurement at
a single point on the sample, inhomogeneities in the sample do not play any role. As has been seen
above, it is even possible to distinguish waves with the same wave-number modulus but travelling in
opposite directions.

In combination with the time-resolution it is possible to resolve the wave-number evolution. This is
shown in Fig.~\ref{fig:fourier} where the measured wave vector distribution for different time intervals
is presented when the front of the spin-wave packet passed the laser spot.

The integrated signal from the interval marked as $D$ in Fig.~\ref{fig:measurement} is presented in
Panel~(I). The observed peak with an experimentally measured wave number of $k_{\rm FMR} = (900 \pm
700)~{\rm m}^{-1}$ corresponds to the thermally excited uniform mode in the sample. Panels~(II)-(IV)
contain the measured wave-vector distributions for three consecutive, $8.7~{\rm ns}$ long time intervals
at the moment when the front of the pulse passed the laser spot and was detected. These time slices are
taken from the interval denoted by $C$ in Fig.~\ref{fig:measurement}. Panel~(V) finally shows the
wave-vector distribution for interval $B$ when the measured signal intensity and the wave-vector
distribution have reached a stable regime. By comparing the panels, the different wave-number
contributions at the front of the pulse due to the dispersion of the spin-wave packet can be
distinguished. It is in qualitative agreement with the phase profile of a linear spin-wave packet, which
exhibits characteristic distortions at the front and end of the pulse \cite{Ser06}.

\section{Conclusion}

In conclusion, we have improved the existing wave-vector resolution used in Brillouin light scattering
experiments by including an electro-optical modulator as a reference to calibrate the position of the
diaphragm hole. The EOM beam makes it possible to determine the position where the in-plane wave vector
vanishes with an accuracy comparable to the accuracy of the actual wave-vector measurement itself. For
experiments conducted in the pulse regime, the proposed method does not increase the measurement time
but even cuts it in half under optimum conditions. The applicability of the EOM-based calibration was
tested by measuring the wave vectors of a propagating packet of dipole-dominated spin-waves for
different bias magnetic fields with time resolution. Comparison with the established theory showed a
good agreement.

\section{acknowledgements}
This work has been financially supported by the Matcor Graduate School of Excellence, the
Graduiertenkolleg 792, and the DFG within the SFB/TRR 49.


%
\end{document}